\documentclass[aps,prd,twocolumn,superscriptaddress,tightenlines,nofootinbib]{revtex4-1}
\usepackage[T1]{fontenc}
\usepackage{microtype}
\usepackage[textsize=scriptsize,backgroundcolor=red!70,linecolor=red]{todonotes}
\usepackage{booktabs}
\usepackage{amsmath}
\usepackage{amsfonts}
\usepackage{amssymb}
\usepackage{hyperref}
\usepackage[all]{hypcap}
\usepackage{paralist}
\usepackage{multirow}
\usepackage{graphicx}
\usepackage{dcolumn}
\usepackage{bm}
\usepackage{epsf}
\usepackage{comment}

\newcommand*{\rom}[1]{\expandafter\@slowromancap\romannumeral #1@}

\def\beq{\begin{equation}}
\def\eeq{\end{equation}}
\def\beqa{\begin{eqnarray}}
\def\eeqa{\end{eqnarray}}

\newcommand{\du}[1]{{\bf\color{red}DU: #1}}

\begin{document}
\DeclareGraphicsExtensions{.eps,.ps}

\title{Searching for neutrino self-interactions at future muon colliders}
	
\author{Hongkai Liu,}
\email[Email Address: ]{hliu6@bnl.gov}
\affiliation{High Energy Theory Group, Physics Department,
		Brookhaven National Laboratory, Upton, New York 11973, USA}
	
\author{Daiki Ueda}
\email[Email Address: ]{daiki.ueda@cern.ch}
\affiliation{Physics Department, Technion – Israel Institute of Technology, Haifa 3200003, Israel}
	
\begin{abstract}
Multi-TeV muon colliders offer a powerful means of accessing new physics coupled to muons while generating clean and intense high-energy neutrino beams via muon decays.
We study a fixed-target experiment leveraging the neutrino beams and a forward detector pointing at the interaction point of the muon collider.
The sensitivity to neutrino self-interactions is analyzed as a feasibility study, focusing on the leptonic scalar $\phi$ exclusively coupled to the Standard Model neutrinos.
Our work shows that projections from both the main and forward detectors can enhance the existing limits by two orders of magnitude, surpassing other future experiments. 


\end{abstract}
	
\maketitle

\section{Introduction}
\label{sec:intor}
Neutrino oscillation, as the only confirmed signal of physics beyond the Standard Model observed in laboratory experiments~\cite{Super-Kamiokande:1998kpq,SNO:2002tuh,KamLAND:2002uet,ParticleDataGroup:2022pth}, indicates the existence of new physics in the neutrino sector. However, non-standard neutrino interactions as candidates for such new physics are highly constrained by searches for reaction processes involving charged leptons. In contrast, active neutrino self-interactions via a neutrinophilic mediator are less constrained and are further well motivated by neutrino masses~\cite{Gelmini:1980re,Chikashige:1980ui,Aulakh:1982yn}, dark matter~\cite{DeGouvea:2019wpf,Kelly:2021mcd}, and Hubble tension~\cite{Lancaster:2017ksf,Kreisch:2019yzn,Blinov:2019gcj,Barenboim:2019tux}.
Currently, experiments associated with large amounts of neutrinos play a vital role in probing the neutrino self-interactions in both the low-energy experiments, such as the decay of charged lepton and meson~\cite{Pasquini:2015fjv,Berryman:2018ogk,Dev:2024twk}, double beta decay~\cite{Burgess:1993xh,Blum:2018ljv,Deppisch:2020sqh}, DUNE~\cite{Kelly:2019wow}, $Z$ invisible decay~\cite{Brdar:2020nbj}, and high-energy colliders~\cite{deGouvea:2019qaz,Kelly:2021mcd,Bai:2024kmt,deLima:2024ohf}.
To enhance sensitivity further, future advancements will necessitate the execution of new experiments that utilize a clean (small uncertainties for the energy spectrum) and high-flux neutrino beam.
In light of this situation, muon collider experiments could significantly contribute to advancing this future initiative.

The muon collider, as a multi-TeV lepton collider, has recently received significant attention~\cite{Delahaye:2019omf, Bartosik:2020xwr,Long:2020wfp,MuonCollider:2022ded,MuonCollider:2022nsa,Accettura:2023ked,InternationalMuonCollider:2024jyv,MuCoL:2024oxj,Hamada:2022mua}.  Notably, positive muons have been successfully accelerated to 100~keV by a radio-frequency cavity~\cite{Aritome:2024jiv}. The muon collider holds great potential for exploring new physics, as it can achieve the same center-of-mass energy as the current LHC while providing a much cleaner experimental environment (see, {\it e.g.}, Refs.~\cite{MuonCollider:2022xlm,Aime:2022flm} for reviews).
The inherent instability of muons is a significant challenge in the design of a muon collider. 
However, this very instability also generates an intense, predictable, and high-energy neutrino flux, making the muon collider a neutrino factory (see, {\it e.g}.,~\cite{DeRujula:1998umv,Barger:1999fs,Cervera:2000kp,Freund:2001ui,NeutrinoFactory:2004odt,Huber:2006wb,Bross:2007ts,Geer:2007kn,Huber:2008yx,Tang:2009wp,FernandezMartinez:2010zza,Dighe:2011pa,Ballett:2012rz,Christensen:2013va,Bogacz:2022xsj,Kitano:2024kdv,Denton:2024glz}). 

In this letter, we consider the construction of a forward detector at the muon collider and perform a feasibility study of the neutrino self-interaction search using the neutrino beam.
As a benchmark model, we examine the neutrinophilic scalar, in which a massive complex scalar $\phi$, carrying lepton number $L=-2$, couples to the SM neutrinos exclusively,
\begin{align}
\mathcal{L}\supset \frac{1}{2}\lambda_{\alpha\beta} \overline{\nu^c}_{\alpha} P_L \nu_{\beta} \phi+{\rm h.c.},\label{eq:int}
\end{align}
where $\alpha $ and $ \beta$ are flavor indices. Please look, {\it e.g.,} at Refs.~\cite{Krnjaic:2017zlz,deGouvea:2019qaz,Berryman:2022hds} for some ultraviolet completions of this coupling. 
The subsequent discussions focus on scenario $\alpha,\beta=\mu$ as a natural choice at muon colliders.  
The forward detector studied in this letter would enable the identification of wrong-sign muons, serving as a smoking-gun signature for searches of the neutrinophilic scalar $\phi$. Meanwhile, the main detector at the interaction point would provide complementary sensitivity, particularly for probing heavy states.
Studying other flavor combinations, such as $\alpha, \beta = e$ and $\alpha = e, \beta = \mu$, would present greater challenges at the muon collider. 
This difficulty arises from the signal suppression due to the $\nu_e$ PDF of the muon beam at the main detector and the challenges associated with the charge identification of electrons in the forward detector.
In the following sections, we will outline the anticipated reach resulting from the integration of the capabilities of these two detectors.

\section{Muon collider and neutrino flux}
\label{sec:MuC}
We outline several critical parameters related to the muon beam conditions to facilitate further analysis.
%
%
Throughout this work, we perform analyses based on tentative parameters presented by the International Muon Collider Collaboration~\cite{InternationalMuonCollider:2024jyv}. The integrated luminosity for our analysis is 1 (10) ab$^{-1}$ at the $\sqrt{s}=3$ (10) TeV muon collider.
The number of muons delivered to the accelerator ring per year is estimated as $
N_{\mu}= 2.8\times 10^{20}\left(\frac{{\rm Muons/bunch}}{1.8\times 10^{12}}\right)\left(\frac{{\rm Repetation~rate}~[{\rm Hz}]}{5~{\rm Hz}}\right)$, which is $3.5\times 10^{20}$ ($2.8\times 10^{20}$) for $\sqrt{s}=3$  (10) TeV.
The muons are unstable and eventually decay into neutrinos $\mu^{\pm}\to e^{\pm} +\nu_e (\bar{\nu}_e)+\bar{\nu}_{\mu}(\nu_{\mu})$, which provide opportunities to search for new physics, {\it e.g.,} neutriophilic particles.
In particular, those decays in a straight section of the collider ring provide highly collimated neutrino beams.
The fraction of muons that decay in the straight section is given by the ratio of the straight section length, $L_{\rm straight}$, to the collider circumference, $C_{\rm coll}$, which is 4.5~(10) km for 3~(10)~TeV muon collider~\cite{InternationalMuonCollider:2024jyv}.
A region of at least $L_{\rm straight}=10$ m without fields is required for the installation of the detector close to the Interaction Point (IP)~\cite{InternationalMuonCollider:2024jyv}, and the number of collimated neutrinos per year is estimated as $N_{\nu}= 1.0\times 10^{-3} N_\mu
\left(
\frac{L_{\rm straight}~[{\rm m}]}{10~{\rm m}}
\right)\left(\frac{10~{\rm km}}{C_{\rm coll}~[{\rm km}]}\right)$, which is $7.7\times 10^{17}$~($2.8\times 10^{17}$) for $\sqrt{s}=3$  (10) TeV.
Depending on the length of the straight section, we could expect a few or 10 times more neutrinos at the practical collider~\cite{InternationalMuonCollider:2024jyv}.

The energy spectrum of the neutrinos from muon decay is given as~\cite{Barger:1999fs} $
d n_{\nu_{\mu}}/dx=2 x^2 \left(3-2x\right)$, and $d n_{\nu_e}/dx= 12 x^2 \left(1-x\right)$, where $x \equiv  E_{\nu}/E_{\mu}$ with the neutrino energy $E_{\nu}$ and the muon beam energy $E_{\mu}$ in the laboratory frame. For unpolarized muons, the angular distribution of the neutrino is isotropic and provided as $dP_{\nu}/d\cos\theta_{\rm rest} =1/2$ in the rest muon frame.
In the laboratory frame, the angular distribution is evaluated as $
d P_{\nu}/d\cos\theta_{\rm lab}=1/\left(2 \gamma^2 \left(1-\beta \cos\theta_{\rm lab}\right)^2\right)$ with $\gamma=E_{\mu}/m_{\mu}$ and $\gamma\beta=p_{\mu}/m_{\mu}$.
This angular distribution is highly collimated to the forward regions\footnote{For the 3 TeV muon collider experiment, 99 \% of the neutrinos are generated in the forward region of $0.7$ mrad.}.
%

\section{Neutrinophilic scalar production}

\begin{figure}
	\centering
	\includegraphics[width=0.48\textwidth]{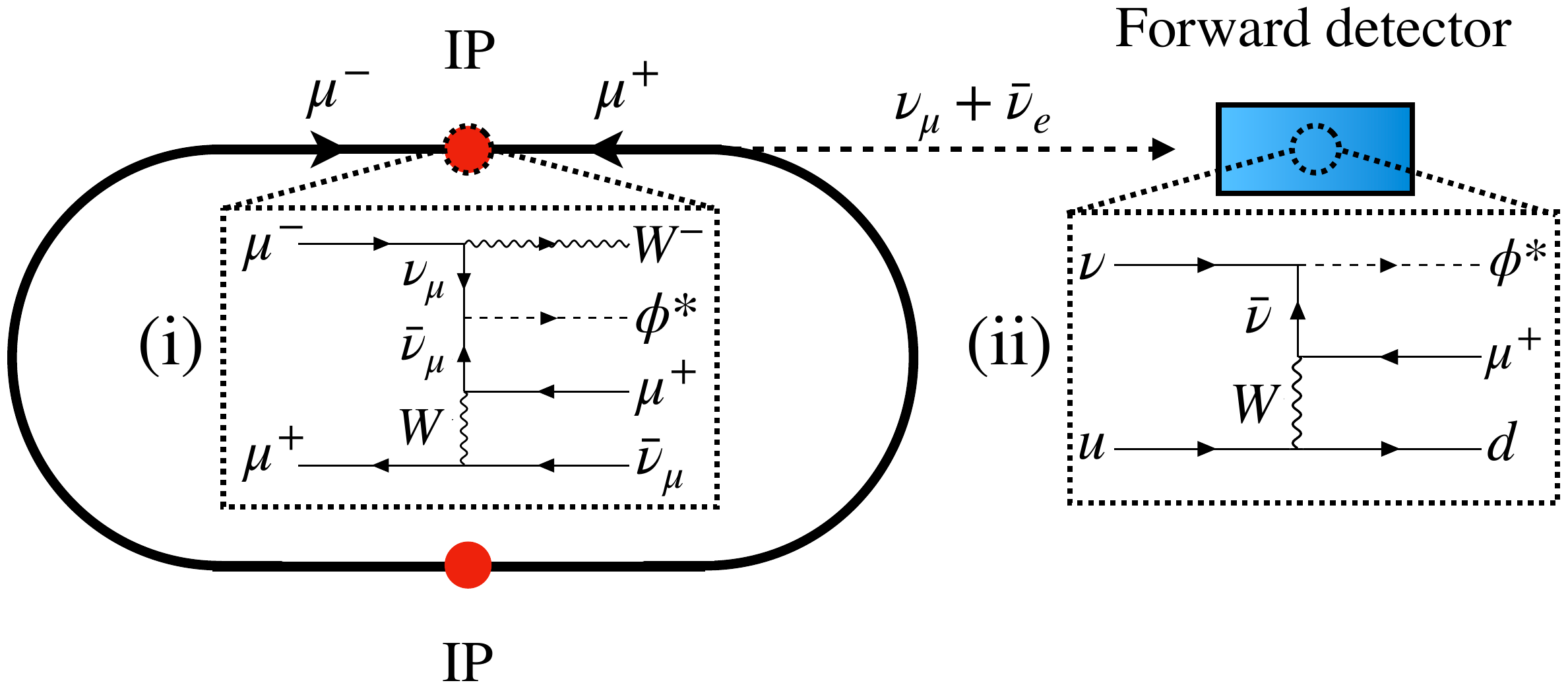}
	\caption{
		The schematic picture of the experiments.
        The scalar $\phi$ may be produced by two scenarios: (i) productions at the IP of the muon collider, as illustrated on the left, and (ii) emissions by high energy neutrinos from muon decay at the forward detector, as shown on the right.
	}
	\label{fig:exp}
\end{figure}

The neutrinophilic scalar $\phi$ may be produced by two different scenarios (see Fig.~\ref{fig:exp}):(i) productions by the muon collisions at the IP (Sec.~\ref{sec:Main_det}) and (ii) radiations from the high-energy neutrinos via the interactions with materials (Sec.~\ref{sec:Forward_det}).
In the first scenario (i), $\phi$ generates extra missing energy in the main detector.
Meanwhile, the second scenario (ii) requires a forward detector to capture the neutrinos in the beam direction.
We shall consider these two scenarios in more detail below.

\subsection{Fixed-target experiment}
\label{sec:Forward_det}
To detect the scalar $\phi$ produced via neutrino bremsstrahlung processes, as shown in the right diagram of Fig.~\ref{fig:exp}, we consider placing far detectors in the forward region of the straight section.
In what follows, we will conduct an analysis based on the performance of the existing detector at the Forward Search Experiment (FASER)~\cite{FASER:2024ykc,FASER:2023zcr}.
Under this setup, a passive tungsten-emulsion neutrino detector (corresponding to FASER$\nu$) with a mass of 1.2 tonnes can be regarded as a fixed-target, allowing us to conduct a fixed-target experiment.
This experiment has a typical energy scale of $\sqrt{2 E_{\nu} m_N}\lesssim 55\text{--}100$ GeV and is highly sensitive to tiny coupling regions of parameter space due to the high luminosity provided by the dense target.

\begin{figure}
	\centering
	\includegraphics[width=.4\textwidth]{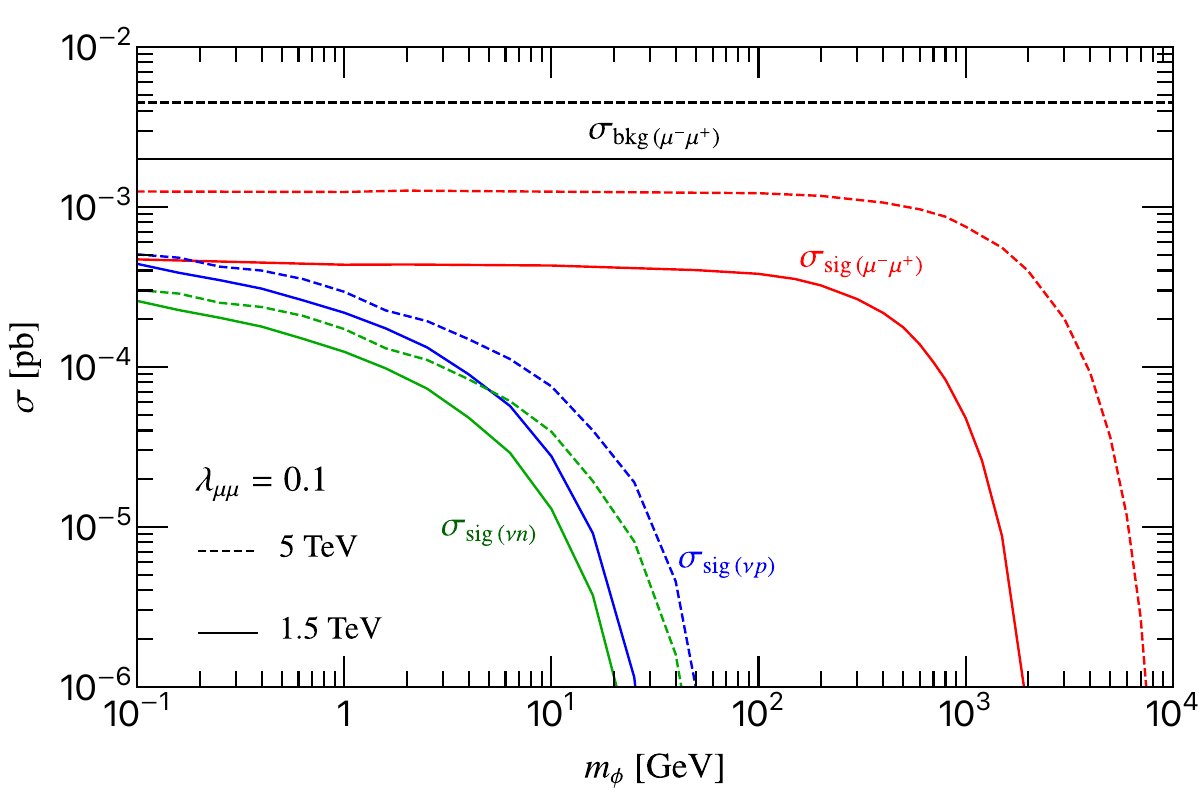}
	\caption{The cross sections of $\phi$ production as a function of $m_\phi$. The blue~(green) curves represent the $\nu p$~($\nu n$) scattering at the forward detector. At the main detector, the signal and background cross sections are shown by red and black curves, respectively. 
    Solid and dashed curves correspond to the 1.5 and 5 TeV incoming neutrino~(muon) energies at the forward (main) detector in the laboratory frame, respectively.
    }
	\label{fig:phi_BD}
\end{figure}

This work focuses on the $\nu_{\mu}$ beam associated with the $\mu^-$ beam and the forward detector optimized for detecting this neutrino beam. 
However, the $\bar{\nu}_e$ beam could also be utilized to search other new physics models through similar analyses with the approach using the $\nu_{\mu}$ beam described below.
The actual location of the forward detector would be determined, {\it e.g.}, based on the construction of the experimental facilities and physical performances\footnote{In our scenario, the resultant sensitivities are slightly improved when focusing on a forward detector along the $\mu^+$ beam direction because antineutrinos interact with down quarks, which are more abundant in a detector with more neutrons than protons.
}.
The number of signal events at the forward detector is estimated as
\begin{align}
&N_{\rm sig}= N_{\nu_{\mu}} \int d E_{\nu}\frac{dn_{\nu_{\mu}}}{dE_{\nu_{\mu}}}\cdot \int  d E_{\mu^+} d \cos\theta_{\mu^+} \nonumber\\
&\quad\quad\quad\quad\sum_{i=n,p}\frac{d^2 \sigma_{\nu_{\mu}+i\to \phi^*+\mu^+ X}}{d E_{\mu^+}d\cos\theta_{\mu^+}}\cdot \rho_{\rm det}^{(i)}\cdot L_{\rm det}\cdot {\rm Acc},\label{eq:sig_BD}
\end{align}
where $E_{\mu^+}$ and $\cos\theta_{\mu^+}$ denote the energy and the production angle of the final state $\mu^+$, $\rho_{\rm det}^{(i)}$ is the number density of protons (neutrons) in the detector for $i=p$ ($n$), $L_{\rm det}=1$ m is the length of the detector, and ${\rm Acc}$ represents an acceptance of the signal detection. $\sigma_{\nu_{\mu}+i\to \phi^*+\mu^+ X}$ is the signal cross section estimated by \texttt{MadGraph5\_aMC@NLO}~\cite{Alwall:2011uj} and \texttt{FeynRules}~\cite{Christensen:2008py}. 
In Fig.~\ref{fig:phi_BD}, we show the total cross section of the signal event as a function of the scalar mass $m_{\phi}$.
The blue (green) curves represent the neutrino-proton~(neutron) interactions.
The solid (dotted) curves correspond to 1.5 TeV (5 TeV) incoming neutrino energy. 
The channel shown in the right diagram of Fig.~\ref{fig:exp} is a clean signature of the scalar $\phi$ production by identifying the charge of the final-state muon.
This is because negative charged muons are produced by the $\nu_{\mu}$ beams via the SM weak interactions but those are reducible under the charge identification of the muons.

To ensure a clear signal, we need to impose an acceptance criterion on the signal events.
Based on the analysis in the FASER, we assume that the acceptance appearing in Eq.~\ref{eq:sig_BD} is provided as ${\rm Acc}=\Theta \left(1~{\rm TeV}-E_\mu\right)\cdot\Theta \left(E_\mu-100~{\rm GeV}\right)\cdot \Theta \left(25~{\rm mrad}-\theta_\mu\right)$
with the Heaviside function $\Theta$.
According to Ref.~\cite{FASER:2024ykc}, muon identification becomes unreliable for energies exceeding 1 TeV.
In addition, to reduce the misidentification rate of the muon, track muon energy is required to fulfill $E_{\mu}\gtrsim100$ GeV~\cite{FASER:2023zcr}.
The angular acceptance, $\theta_\mu <25$ mrad, comes from the detector size of electronic detector components of FASER~\cite{FASER:2023zcr,FASER:2024ref}. 
Kinematical distributions relevant to the acceptance are shown in the Supplemental Material in Fig.~\ref{fig:kine_BD}.
According to these distributions, we find that most of the $\mu^+$ energy is carried by a momentum along the incoming neutrino beam direction, and a peak of its momentum is around 200\text{--}500 GeV.
In particular, the peak of the $\mu^+$ energy is broader for higher muon beam energy.
For instance, the efficiencies of the 3 TeV muon collider experiment for the energy, angular, and combined cuts are 0.87 (0.80), 0.70 (0.60), and 0.67 (0.56), respectively, for $m_{\phi}=0.1$ (10) GeV.
Also, the efficiencies of the 10 TeV muon collider experiment for the energy, angular, and combined cuts are 0.45 (0.53), 0.90 (0.87), and 0.38 (0.43), respectively, for $m_{\phi}=0.1$ (10) GeV.
%
%
%

We next address potential background (BG) events, which are classified into three types, as follows: (a) Neutrino beam-induced BG --- For $\nu_{\mu}$ beam, oppositely charged muons against the signal event are produced via the SM-charged current processes, {\it e.g.}, deep inelastic scattering.
	The detector with a magnetic field can identify the muon charges accurately for their energy below 1 TeV.
	Thus, $\nu_{\mu}$ induced the charged current processes are reducible BG events.
	In addition, neutrino beams may produce charged particles, potentially mimicking charged muon.
    However, this type of BG would be much reduced because, in FASER, scintillator stations and tracking systems enable the identification of charged particles passing through the entire length of FASER with inefficiencies smaller than $10^{-7}$~\cite{FASER:2023zcr}.
    As studied in Ref.~\cite{Adhikary:2024tvl}, $\mu^+$ coming from charm hadron decays could also be significantly reduced.
    (b) Muon beam-induced BG --- The neutrinos can be produced at the IP by $\mu^+ \mu^-$ collisions.
	However, compared with the neutrino beams from the muon decays, the cross section of the weak interaction processes suppresses its flux.
	For instance, the total cross section of a process $\mu^+ +\mu^- \to \nu_{\mu}+\bar{\nu}_{\mu}$ is $54$ pb for the 10 TeV muon collider experiment, and the number of produced $\nu_{\mu}$ or $\bar{\nu}_{\mu}$ is $5.4\times 10^8$ for 10 ab$^{-1}$.
	Although these $\bar{\nu}_{\mu}$ cannot reach the forward detector because these point to the opposite direction against $\mu^-$ beam direction, the number of produced $\mu^-$ at the forward detector by $\nu_{\mu}$ of this process is $\sim \mathcal{O}(10)$.
    The $\bar{\nu}_{\mu}$ produced through other processes are expected to be further suppressed, rendering the BG induced by these $\bar{\nu}_{\mu}$ negligible.
	%
	%
	%
	%
    (c) Beam-unrelated BG --- The BG events not associated with the beam is primarily attributed to cosmic rays.
	However, the cosmic muon fluxes can decrease by the detector's depth below the ground surface.
	%
	%
	%
	%
	In addition, most cosmic muons can be removed by the angular cuts.
	%
	%
	By putting cosmic-muon veto, cosmic-muon BG can be further reducible.
	%
    %
    Although the above estimates suggest that these types of BG events can be much reduced, we defer further quantitative analysis to the phase of detector design dedicated to the muon collider.
	In Fig.~\ref{fig:bound}, we assume 10 and 100 signal events to estimate the expected reach, as shown by the blue and light blue curves, respectively.

\subsection{$\mu^+\mu^-$ collider experiment}
\label{sec:Main_det} 
%
%
The main detector at the interaction point offers a complementary approach with the fixed-target experiment, enabling the exploration of heavier $\phi$. The new scalar may be produced via the hard scattering $\mu^-\mu^+\to W^-\mu^+\bar\nu_\mu\phi^*$ as shown in the left diagram of Fig.~\ref{fig:exp}.
We require the $W$ boson to decay into di-jet for two reasons. 
First, $W\to jj$ channel has the largest branching ratio, and jets are less common at the lepton collider, so they do not suffer from considerable BG contamination. Second, the BG contributions from the SM 2-to-3 process $\mu^-\mu^+\to W^- \mu^+ \nu_\mu$ can be significantly reduced because of $p_{\rm miss}^2 = p_{\nu}^2 = 0$ once we fully reconstruct the momentum of the signal missing energy, {\it i.e.,} $p_{\rm miss}^2 = (p_\phi + p_{\bar\nu})^2 > m_\phi^2$. 
According to these, the leading SM contributions are from the process $\mu^-\mu^+\to W^- \mu^+ \nu_\mu \nu\bar\nu$.
We use \texttt{MadGraph5\_aMC@NLO} to calculate the signal and BG cross sections with some basic cuts applied: $p_{T,j},p_{T,\ell} > 50$~GeV, $E_{T,\text{miss}} > 80$~GeV, and $\mid\eta_{j}\mid,\mid\eta_{\ell}\mid < 3.0$, where $p_{T,j~(\ell)}$ and $\eta_{j~(\ell)}$ denote the transverse momentum and pseudorapidity of jets (charged leptons), respectively, and  $E_{T,\text{miss}}$ is the missing transverse energy. 
The red and black curves in Fig.~\ref{fig:phi_BD} show the resulting signal and BG cross sections, respectively.

\begin{figure}
	\centering
	\includegraphics[width=0.48\textwidth]{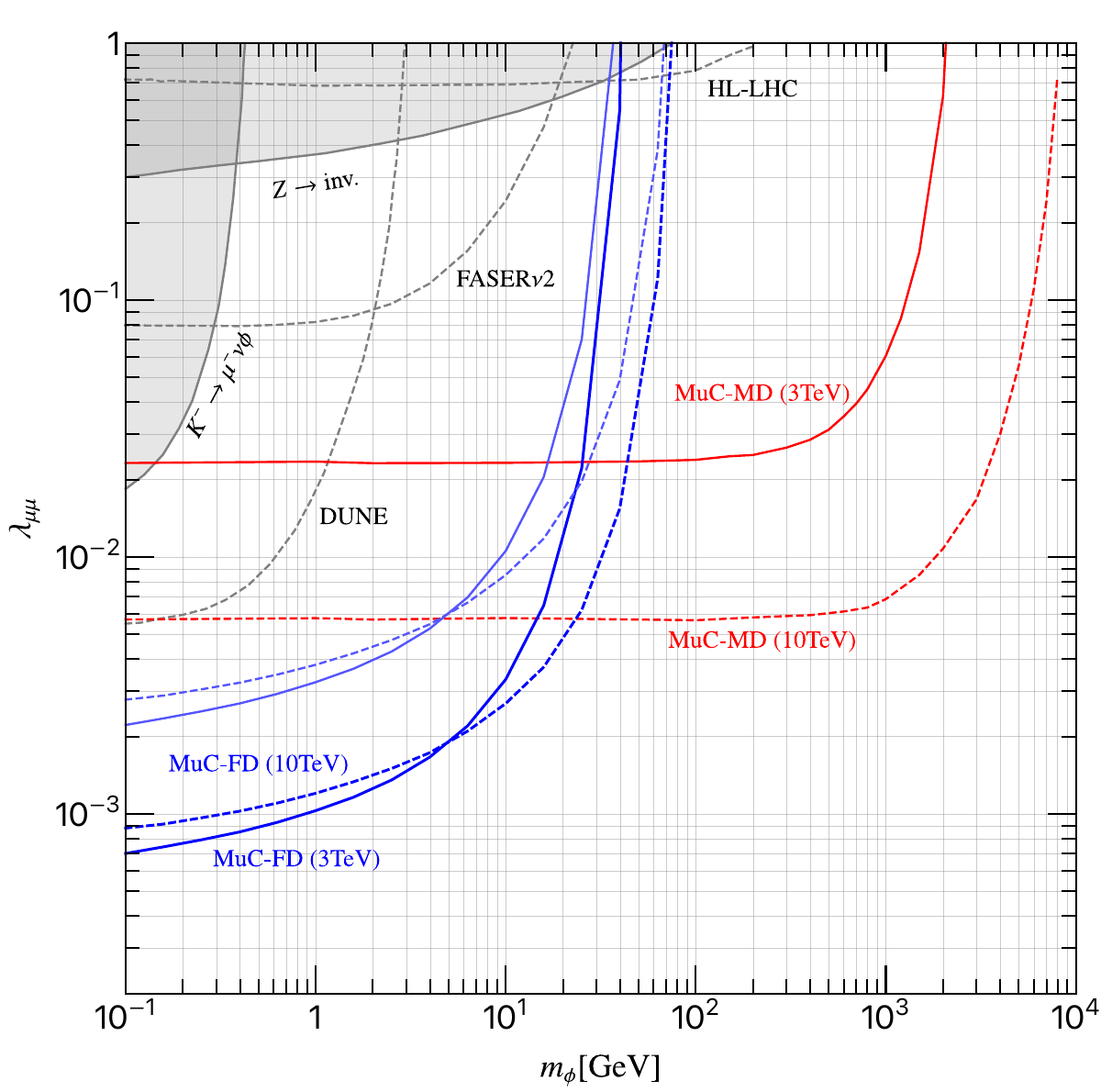}
	\caption{The bounds on the neutrinophilic scalar.
    The colored curves represent the results from the main (red) and forward detectors (blue and light blue), respectively.
    The projections from the forward detector are set by requiring 10 (blue) and 100 (light blue) signal events. While, for the main detector, we show $2\sigma$ bounds. 
    Colored solid~(dashed) curves represent the results of a 3~(10) TeV muon collider.
    The current bounds from meson~\cite{Berryman:2018ogk,Kelly:2021mcd,Dev:2024twk} and $Z$~\cite{Brdar:2020nbj} decay are shaded by gray. Future projections from HL-LHC~\cite{deGouvea:2019qaz}, FASER$\nu$2~\cite{Bai:2024kmt}, and DUNE~\cite{Kelly:2019wow} are indicated by dashed gray lines.}
	\label{fig:bound}
\end{figure}

Some important kinematics distributions at 10~TeV muon collider are shown in the Supplemental Material in Fig.~\ref{fig:kine}.
In particular, the invariant mass $M_{\rm miss} = \sqrt{p^2_{\rm miss}}$ and the pseudorapidity $\eta_{\rm miss}$ of the missing momentum can provide crucial information for distinguishing the signal from the BG. 
Based on the kinematic features, we further require $-0.5~(0)<\eta_\ell<2.7$, $p_{T,\ell} > 150~(200)~\text{GeV}$, $\mid\eta_{\rm miss}\mid < 1.5$, and $0.5~(2)~\text{TeV}+m_\phi/2 < M_{\rm miss} < 2~(8)~\text{TeV}+m_\phi/4$ for $\sqrt{s} = $3~(10) TeV 
to increase the significance $S/\sqrt{B}$, where $S~(B)$ is the signal~(BG) event number after the cuts. After combining the other signal channel $\mu^-\mu^+\to W^+\mu^-\nu_\mu\phi$, we show the 2~$\sigma$ bounds with the red curves in Fig.~\ref{fig:bound}.

\section{Summary}
\label{sec:sum} 
In this letter, we investigated the prospects of detecting the neutrino self-interactions at the muon collider experiment, focusing on the massive complex scalar interacting with the SM muon neutrinos exclusively.
Previous research on muon collider experiments has primarily concentrated on identifying new particle signatures at the IP.
In contrast, we explored the feasibility of detecting new particles utilizing forward detectors in conjunction with detection at the IP. 
In the muon collider experiment, the muon beams are unstable, and collimated high-intensity neutrino beams are produced in their straight section.
These neutrino beams traverse considerable distances (of order 100 m from the IP) to the forward detector and potentially produce the new particles at the forward detector through the interaction of the neutrino beams with the detector.
The neutrinophilic scalar $\phi$ 
generates a wrong-sign muon as a distinctive signal, which can be accurately identified for muon energy below 1 TeV.
Regarding the signal at the IP of the muon collider, the scalar can be produced in the process $\mu^-\mu^+\to W^-\mu^+\bar\nu_\mu\phi^*$ with higher center-of-mass energy.
Complementing the fixed-target experiment, the searches at the main detector potentially explore scalar masses reaching several TeV.
The results of the sensitivity analysis for the model are illustrated in Fig.~\ref{fig:bound}.
Our analysis indicates that the muon collider experiment, as examined in the reference model, is sensitive to parameter space regions that extend significantly beyond the current constraints and future experiments, such as FASER$\nu2$, DUNE, and HL-LHC.

\section*{Note added}
While completing this work, we noticed that Ref.~\cite{Adhikary:2024tvl} was posted on the arXiv.

\begin{acknowledgments}
We would like to thank Hooman Davoudiasl, Peter Denton, and Yotam Soreq for their useful comments on the manuscript.
HL is supported by the U.S. Department of Energy under Grant Contract DE-SC0012704.
DU is supposed by grants from the ISF (No.~1002/23 and 597/24) and the BSF (No.~2021800).
\end{acknowledgments}


\twocolumngrid
\vspace{-8pt}
\section*{References}
\vspace{-10pt}
\def\bibsection{}
\bibliography{MC}

\clearpage
\onecolumngrid

\setcounter{equation}{0}
\setcounter{figure}{0}
\setcounter{table}{0}
\setcounter{section}{0}
\makeatletter
\renewcommand{\theequation}{S\arabic{equation}}
\renewcommand{\thefigure}{S\arabic{figure}}
\renewcommand{\thetable}{S\arabic{table}}
\renewcommand{\thesection}{S\arabic{section}}
\begin{center}
 \large{\bf 
 Searching for neutrino self-interactions at future muon collider
 }\\
 Supplemental Material
\end{center}
\begin{center}
Hongkai Liu, and 
Daiki Ueda
\end{center}

In the supplemental material, we further detail kinematical distributions at the fixed-target experiment and $\mu^+\mu^-$ collider experiment.

\section{Kinematical distributions at fixed-target experiment}
\label{sec:app:kin_BD}
\begin{figure}
	\centering
    \includegraphics[width=0.4\textwidth]{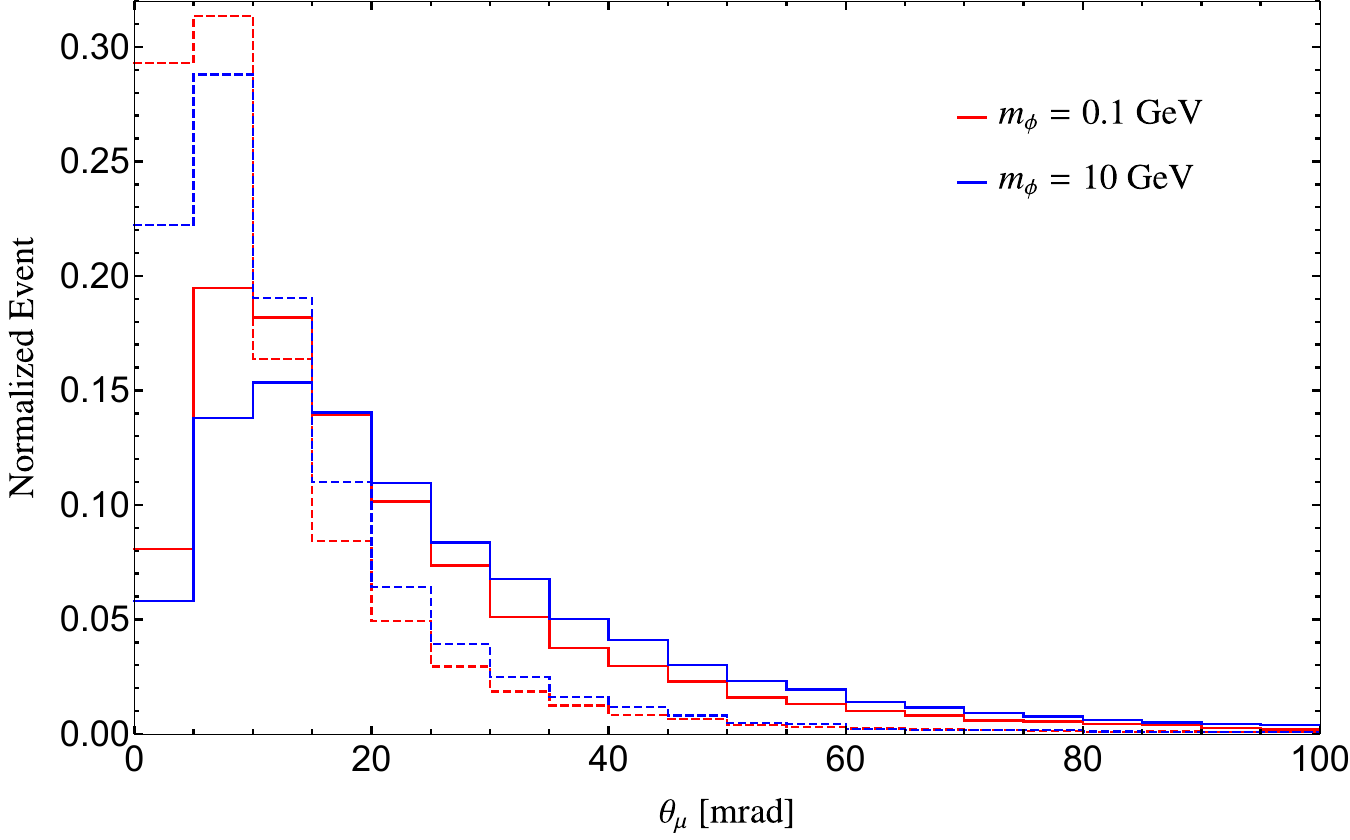}
	\includegraphics[width=0.4\textwidth]{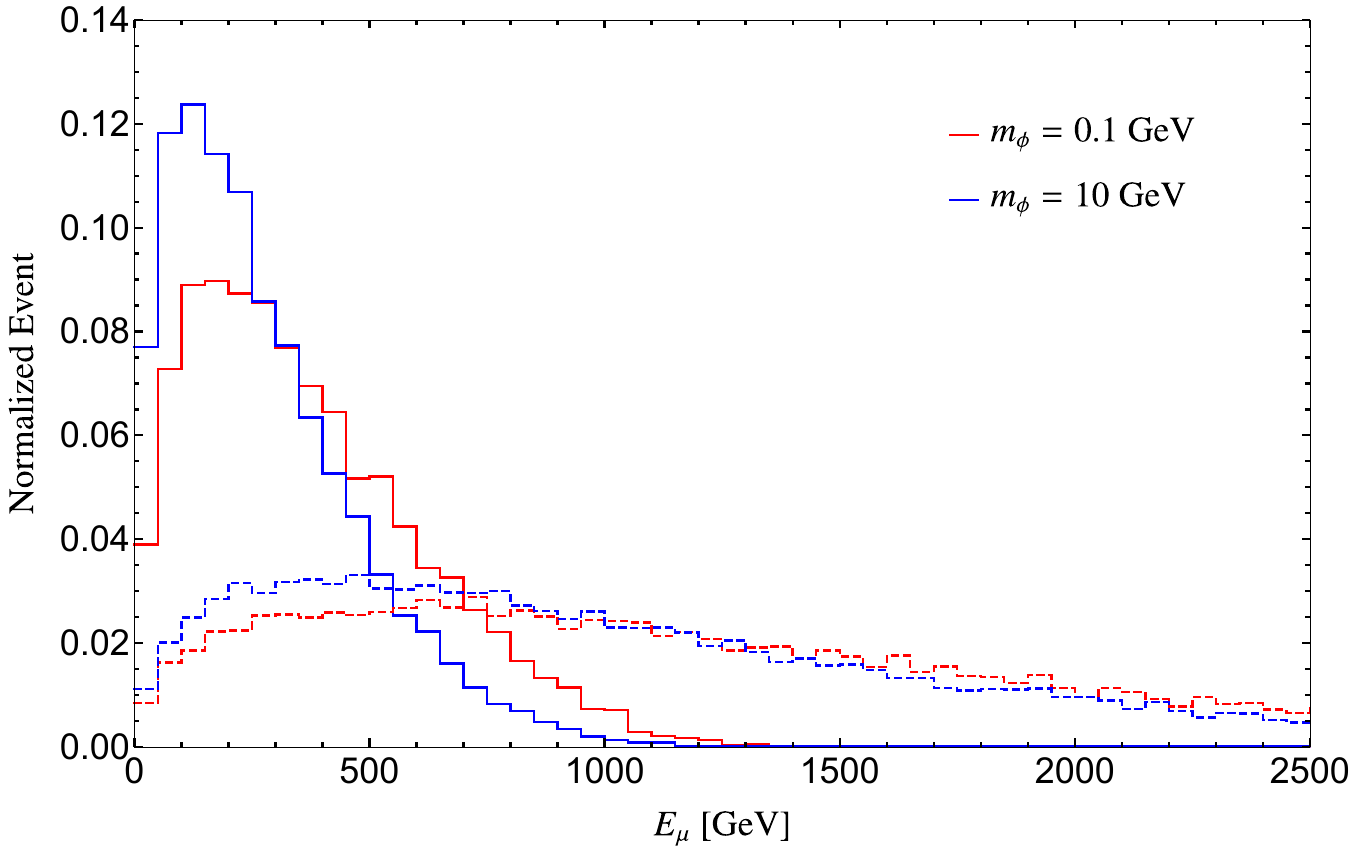}
	\caption{The distributions of final-state $\mu^+$ angle  $\theta_{\mu}$ (left), and  energy $E_{\mu}$ (right) at the forward detector for $m_{\phi}=0.1$ GeV (red) and $m_{\phi}=10$ GeV (blue).
    Solid and dashed curves correspond to 1.5 and 5 TeV incoming muon energies in the laboratory frame. 
	}
	\label{fig:kine_BD}
\end{figure}

Kinematic distributions related to the acceptance at the fixed-target experiment are shown in Fig.~\ref{fig:kine_BD}. 
The left and right panels represent the distributions of the final state $\mu^+$ angle $\theta_{\mu}$ and energy $E_{\mu}$ for the signal event at the forward detector.
The red and blue curves denote $m_{\phi}=0.1$ GeV and $m_{\phi}=10$ GeV.
The solid and dashed curves correspond to the 3 and 10 TeV muon collider experiments.
From the left panel of Fig.~\ref{fig:kine_BD}, the final state $\mu^+$ momentum is almost carried by its beam direction component, and it is more collimated along the neutrino beam axis for higher muon beam energy.
The right panel of Fig.~\ref{fig:kine_BD} shows that the final state $\mu^+$ energy peaks in around 200-500 GeV and implies that enough signal events remain after imposing cuts.
In particular, this peak becomes broader for higher muon beam energy, and the energy cut effect on the signal event becomes significant.

\section{Kinematical distributions at $\mu^+\mu^-$ collider experiment}
\label{sec:app:kin_mu+mu-}
\begin{figure}
	\centering
    \includegraphics[width=0.4\textwidth]{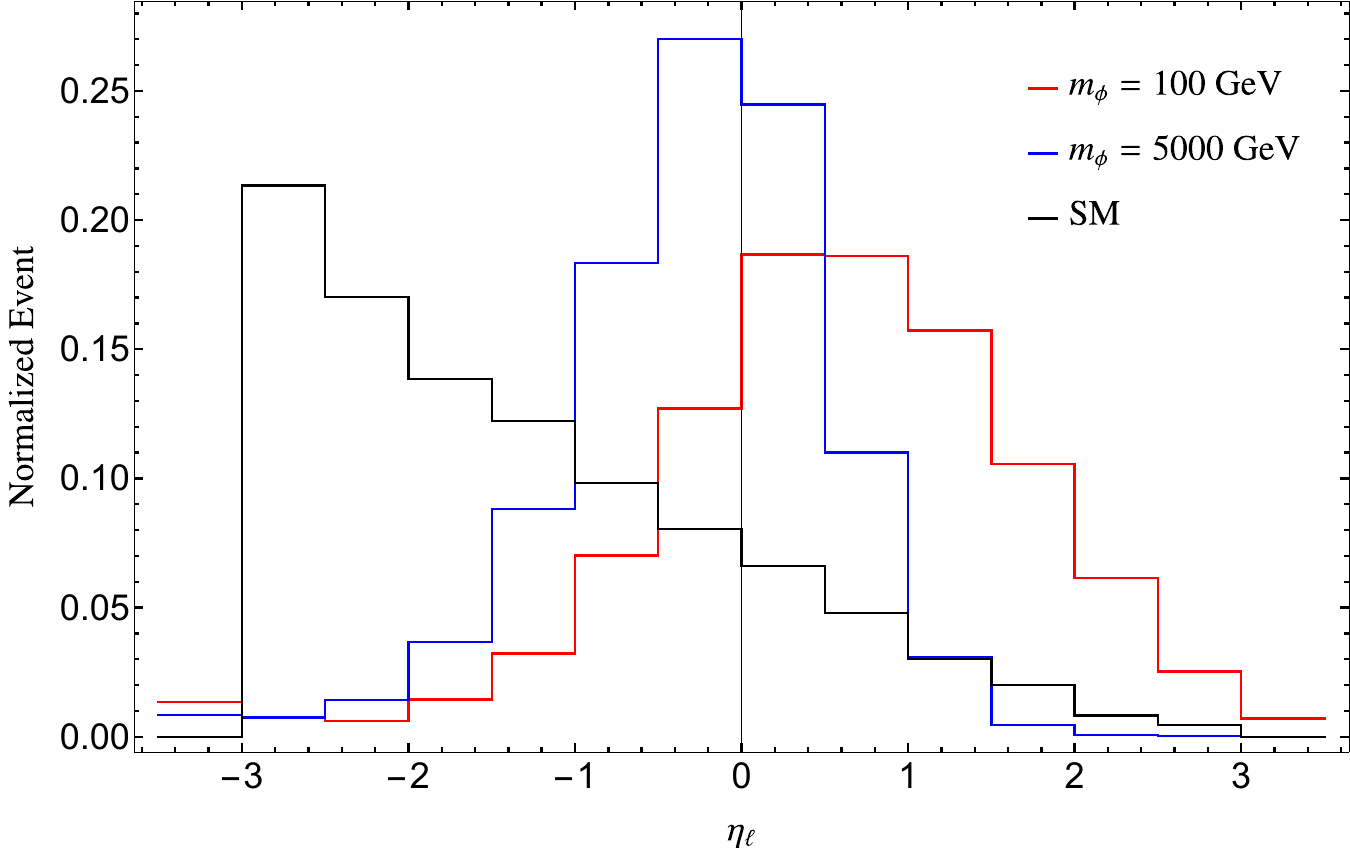}
	\includegraphics[width=0.4\textwidth]{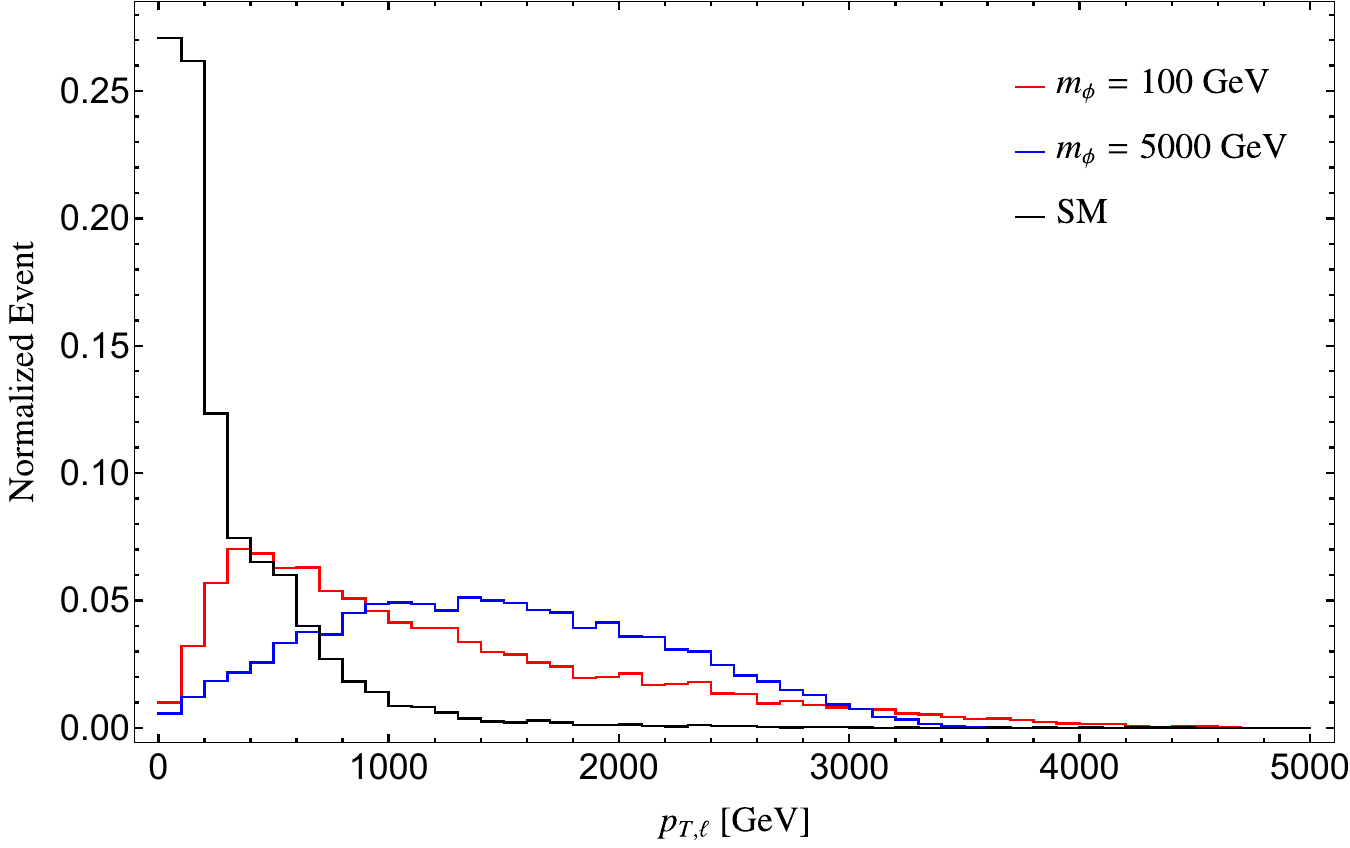}
	\includegraphics[width=0.4\textwidth]{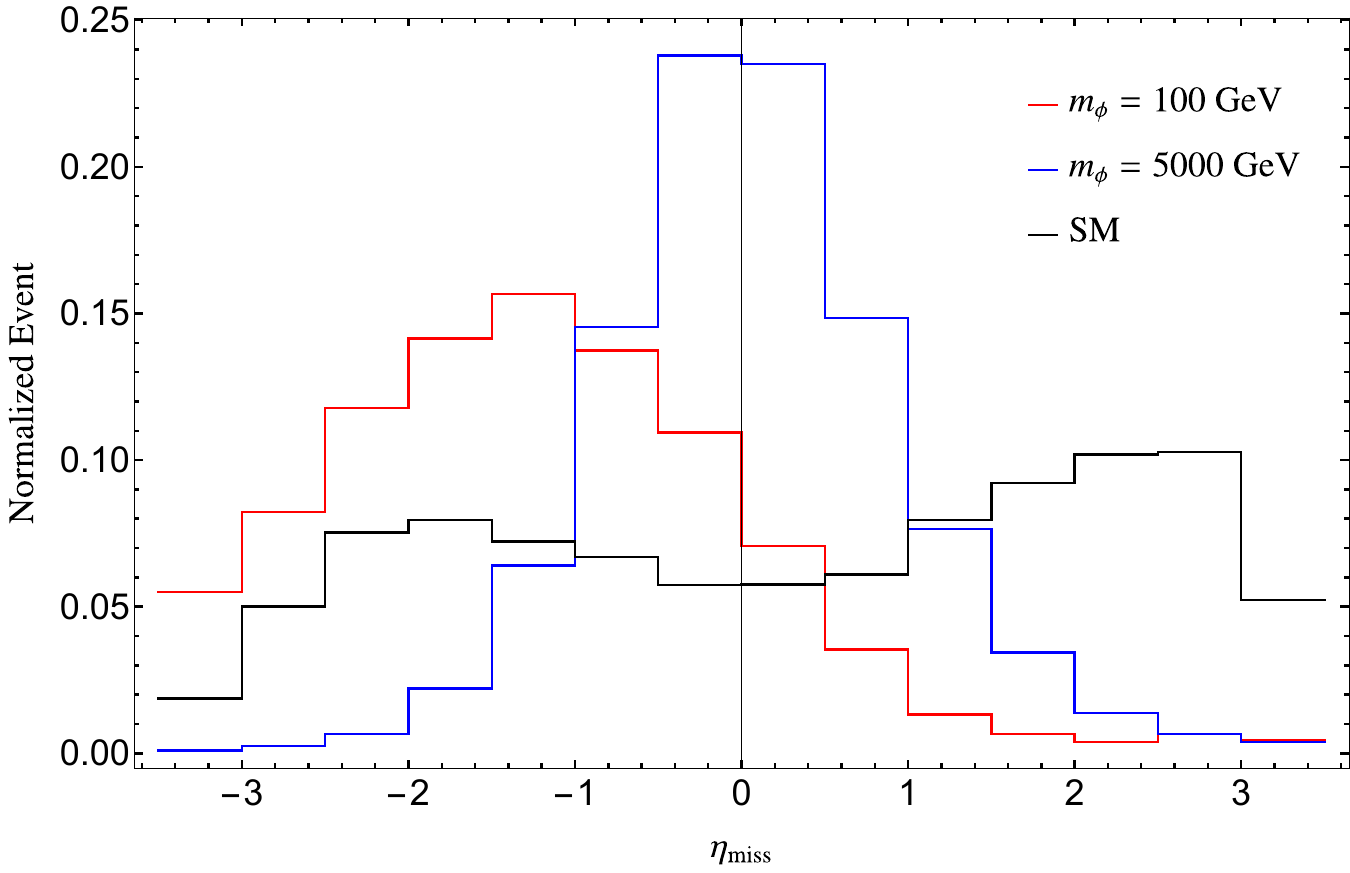}
	\includegraphics[width=0.4\textwidth]{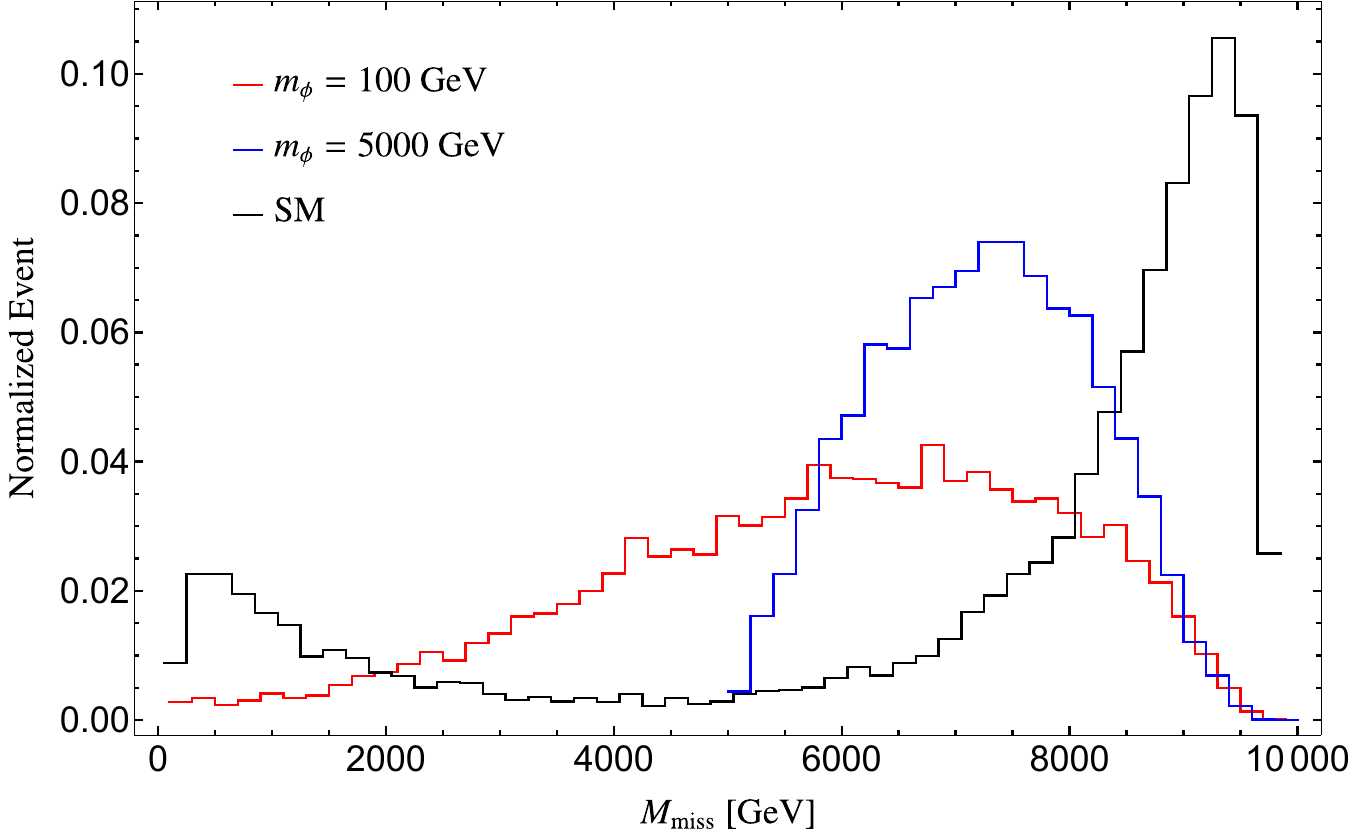}
	\caption{The distributions of $\eta_\ell$ (upper left), $p_{T,\ell}$ (upper right), $\eta_{\rm miss}$ (lower left), and $M_{\rm miss}$ (lower right) for signal (colored curves) and BG (black curves) at 10~TeV muon collider.
	}
	\label{fig:kine}
\end{figure}

Some important kinematics distributions at 10~TeV muon collider are shown in Fig.~\ref{fig:kine}.
These features hold even for the 3 TeV muon collider. 
The red~(blue) curves represent the distributions of signal with $m_\phi = 100~(5000)$~GeV. The black curves show the SM BG.
The upper right panel of  Fig.~\ref{fig:kine} shows the distributions of the antimuon's pseudorapidity $\eta_\ell$.
These distributions indicate that final-state $\mu^+$ in the signal event tends to follow the direction of the initial $\mu^-$, especially for a light $\phi$.
In contrast, for the SM BG, the final-state $\mu^+$ tends to retain the direction of the initial $\mu^+$. 
The $\mu^+$ has higher transverse momentum in the signal compared to the BG, as shown in the upper right panel of Fig.~\ref{fig:kine}.
The lower left panel of Fig.~\ref{fig:kine} shows the distributions of $\eta_{\rm miss}$. 
The missing momentum shifts along the negative-z direction (direction of the initial $\mu^+$ beam) for light $\phi$.
However, for heavy $\phi$, $p_{\rm miss}$ is closer to $p_\phi$ and more central.
In the SM BG, $\eta_{\rm miss}$ has a broader distributions. 
Finally, the invariant mass of missing momentum $M_{\rm miss} = \sqrt{p^2_{\rm miss}}$ can provide crucial information to distinguish signal and BG, as shown by the lower right panel of Fig.~\ref{fig:kine}. 
%

\end{document}